\documentclass[letterpaper, 10 pt, conference]{ieeeconf}  

\IEEEoverridecommandlockouts                              

\usepackage{graphics} 
\usepackage{epsfig} 

\usepackage{romannum}
\usepackage{array}
\usepackage{multirow}
\usepackage{amsmath}
\usepackage{amssymb}
\usepackage{caption}
\usepackage{subcaption}

\renewcommand{\baselinestretch}{0.90}

\title{\LARGE \bf
A Weak Monotonicity Based Muscle Fatigue Detection Algorithm for a Short-Duration Poor Posture Using sEMG Measurements
}

\author{Xinliang Guo, Lei Lu, Mark Robinson, Ying Tan, Kusal Goonewardena and Denny Oetomo
\thanks{X. Guo, M. Robinson, Y. Tan, and D. Oetomo are with the Faculty of Engineering and Information Technology,
        The University of Melbourne, Parkville, VIC 3010, Australia (Email: xinliangg@student.unimelb.edu.au, mrobinson2@student.unimelb.edu.au, yingt@unimelb.edu.au, doetomo@unimelb.edu.au)}
\thanks{L. Lu is with the Institute of Biomedical Engineering, Department of Engineering Science, 
        University of Oxford, Oxford, OX3 7DQ, UK (Email: lei.lu@eng.ox.ac.uk)}
\thanks{K. Goonewardena is with the Elite Akademy Sports Medicine,
        Parkville, VIC 3010, Australia (Email: kusal@eliteakademy.com)}%
}

\begin{document}

\maketitle
\thispagestyle{empty}
\pagestyle{empty}

\begin{abstract}
Muscle fatigue is usually defined as a decrease in the ability to produce force.
The surface electromyography (sEMG) signals have been widely used to provide information about muscle activities including detecting muscle fatigue by various data-driven techniques such as machine learning and statistical approaches.
However, it is well-known that sEMG signals are weak signals (low amplitude of the signals) with a low signal-to-noise ratio, data-driven techniques cannot work well when the quality of the data is poor. In particular, the existing methods are unable to detect muscle fatigue coming from static poses.
This work exploits the concept of weak monotonicity, which has been observed in the process of fatigue, to robustly detect muscle fatigue in the presence of   measurement noises and human variations.
Such a population trend methodology has shown its potential in 
%
%
muscle fatigue detection as demonstrated by the experiment of a static pose.

\end{abstract}

\section{INTRODUCTION}

Muscle fatigue is the failure of a muscle to generate expected force \cite{Cifrek2009}. Normally, it can be a result of vigorous exercise, exhaustive labor, or holding static postures for a prolonged period. 
Fatigue is also one of the most common causes of chronic pain \cite{Roy1989}. Every year, millions of people worldwide suffer from chronic pain. Hence, a system that detects muscle fatigue followed by an appropriate intervention can greatly reduce the risk of chronic pain.

Surface electromyography (sEMG) can be used to detect muscle fatigue \cite{Toro2019}. It has been observed that the frequency spectrum of sEMG signal will downshift when a muscle becomes fatigued \cite{Chowdhury2015}. 
Many signal processing techniques have been proposed to detect muscle fatigue, for example statistical methods \cite{Chowdhury2013} and unsupervised machine learning \cite{R.2012}. It is not surprising that the performance of signal processing depends on the quality of signals. Due to the low signal-to-noise ratio (SNR) of sEMG signals \cite{DeFreitas2012} and large human variations, the existing methods can easily detect the muscle fatigue coming from dynamic movements when the subjects already feel exhausted in the experiments. However, the existing techniques cannot work well in detecting muscle fatigue from static poor posture. It is reported that poor posture presents emerging health risks 
\cite{Ahmad2018}.

%
Although, recent investigations in \cite{Lu2020} have shown that upper-back muscles will indeed fatigue even after sitting with poor posture for only 15 minutes by statistically analyzing experimental results of a population, there is still no systematic way to capture muscle fatigue from static poses using sEMG signals for each individual.  

It has been observed that the downshift of the median frequency of sEMG signals can be used to detect muscle fatigue 
\cite{Jankovic2010}. It suggests that a muscle fatigue has a  a population trend, which can be used in muscle fatigue detection,
and lead to the threshold-based detection algorithm \cite{xie2010fuzzy}.
%
%
Due to the low SNR properties of sEMG and the existence huge human variations, the population trend might not be always dominant or clearly to be observed, leading to the concept of weak monotonicity (WM) \cite{Lu2020a}.
%
This work utilizes the concept of WM to develop a novel fatigue detection algorithm to capture upper-back muscle fatigue during a short-duration poor posture. In order to validate the effectiveness of the proposed algorithm, two set of experiments have been conducted. Experiment 1 collected sEMG signals when subjects were required to sit in a poor posture for 15 minutes while Experiment 2 worked with clinicians to detect muscle fatigue when subjects were sitting in the same poor posture. The data collected in the Experiment 1 was used to tune the parameters of proposed WM-based muscle fatigue detection while Experiment 2  was used to demonstrate the effectiveness of the proposed method.

\section{Methods}

Detecting muscle fatigue coming from long time static poses using sEMG signals is difficult, because of the existence of noise and human variations. Hence, traditional statistical methods are not sensitive enough to capture muscle fatigue for each individual. 
Instead of only using measured data, the population trend observed in muscle fatigue processes can be employed to detect muscle fatigue.
This work focuses on weak monotonicity, which is a kind of robust population trend in the presence of measurement noises and human variations. 

For one measured sEMG signal of a subject during static sitting at time $t$, it is denoted as $x(t)$. The signal will be pre-processed (e.g. filtering, de-trending) and segmented as:
%
\begin{equation}  
s_{T_j}(t_i)=\varphi(t_i, x_{T_j}(t_i)), \ \ \forall t\in [t_{i},t_{i+1}),
\end{equation}
where $t_i=T\times i$ is the $i^{th}$ sampling instant, $T$ is the sampling time, and $[T_{j-1}, T_j], j=1, 2, \cdots, N$, is the time interval of the $j^{th}$ data segment with starting time of $T_{j-1}$ and ending time of $T_j$.  Here $\varphi(\cdot)$ represents an operator to pre-process the sEMG signal. 

This leads to a set of features that can be derived from each data segment. As demonstrated in \cite{Cifrek2009}, we are particularly interested in frequency domain features as the population trend is identified in frequency domain.
\begin{equation}  
F(T_j)= \phi (f, \mathcal{F} (s_{T_j}(t_i))),   \ \ f_0 \leq    f  \leq f_s,
\end{equation}
where $\mathcal{F}(\cdot)$ is an operator to transform a continuous-time signal to a frequency domain signal, $f$ is the instantaneous frequency, and $[f_0, f_s]$ is the frequency range of interests. In the context of sEMG signals, $f_0$ can take $10\sim20$ Hz \cite{Moon2014}, and $f_s$ is around $150$ Hz 
\cite{Phinyomark2018}. The operator $\phi(\cdot)$ maps a frequency-dependent signal of the segmentation to a frequency-invariant feature value. For example, it can represent the calculation of the mean or median values from frequency-domain signal.

With calculating features for each data segment, the corresponding point for the $j^{th}$ data segment $J(T_j)= \{ F(T_j) \}$ is obtained  in order to compute the trend to indicate muscle fatigue.  Traditionally, a threshold is used to detect the muscle fatigue, that is, 
\begin{equation}  
 \sigma(T_j, J(T_j),\theta)\in \{0,1\} \label{trigger}
\end{equation}
where $\sigma(\cdot,\cdot,\cdot)$ is an operator to calculate the threshold of the trajectory where $\theta$ is the threshold. For example, when the downshift of the median frequency of sEMG signals is larger than a given threshold $\theta$, $\sigma(T_j, J(T_j),\theta)$ becomes ``$1$'', indicating the trigger of muscle fatigue. When it is smaller than the threshold $\theta$, $\sigma(T_j, J(T_j),\theta)$ becomes ``$0$''.

However, the trajectory of $J(T_j)$ may be affected by measurement noise and variations, making it less sensitive to detect the muscle fatigue coming from static poses.  

Next, the weak monotonicity (WM) is used to characterize the robust trend. 
For the calculated trajectory $F(T_j)$, the WM is defined if the following inequality is satisfied \cite{Lu2020a},
\begin{equation}  
        \begin{aligned}
    &F(T_{j}) \leq F(T_{j-1})+\delta(T_{j-1}),
    \end{aligned} 
\label{Weak monotonicity}
\end{equation}
where $|\delta(T_{j-1})| \leq \Delta$. Here $\delta(T_{j-1})$ represents the fluctuation of trend due to measurement noises and human variation and $\Delta$ is the WM bound determined by a user-defined variation rate $\Delta_r$, which links to the variation of the data. 

With the definition of WM using inequality (\ref{Weak monotonicity}), for the trajectory with a decreasing trend, the positive and negative sets of the the trajectory can be calculated as follows,
\begin{equation}  
           \left\{  
           \begin{aligned}
           &\mathcal{D}_{Wm^-}=\{F(T_j) \ \big| \  F(T_j) \leq F(T_{j-1}) + \delta (T_{j-1}),\\
           & \qquad \qquad  \qquad  \quad \quad \ \forall \; T_j > T_{j-1} \},\\
            &\mathcal{D}_{Wm^+}=\{F(T_j) \ \big| \  F(T_j) \notin \mathcal{D}_{Wm^-}\},\\
          \end{aligned} \right. 
\end{equation}
Then, the number of data points in each of the two datasets $\mathcal{D}_{Wm^+}$ and $\mathcal{D}_{Wm^-}$ can be calculated as,
\begin{equation}  
             \left\{  
            \begin{aligned}
             &s_{Wm^+} = Cardi(\mathcal{D}_{Wm^+}),\\
             &s_{Wm^-} = Cardi(\mathcal{D}_{Wm^-}),\\
            \end{aligned} \right. \label{the number set}
\end{equation}
where $Cardi(\cdot)$ is an operator to calculate the number of data points in the dataset, $s_{Wm^+}$ and $s_{Wm^-}$ are the number of elements in the two sets. The WM value of a  $F(T_j)$ can be calculated as: 
\begin{equation}  
WM=\frac{s_{Wm+}}{n-1}-\frac{s_{Wm-}}{n-1},
\end{equation}
where $s_{Wm+}$ is the number of positive increasing data points, $s_{Wm-}$ is the number of negative decreasing data points, and $n$ is the total number of data points in the dataset. 

The final WM value is between $-1$ and $1$, where WM value close to $-1$ means the trend is more monotone decreasing, and WM value close to $0$ means the trend is less apparent. Similar to (\ref{trigger}), we can use an appropriate threshold to detect muscle fatigue.

As a special case, the following algorithm is used to design $\sigma(\cdot,\cdot,\cdot)$:
\begin{equation}  
    \left\{  
    \begin{aligned}
    & F(T_{j})(1-\Delta_r) \leq F_{int}-F_{th}, \\
    & WM(T_{j}) \leq WM_{th}, \\
    \end{aligned} 
    \right., \label{WM-based algorithm}
\end{equation}
where $\Delta_r$ is the WM bound variation rate, $F_{int}$ is the frequency at the beginning of the trajectory, $F_{th}$ is the frequency shift threshold, $WM(T_{j})$ is the WM value obtained at the same time point, and $WM_{th}$ is the WM value threshold. 

The frequency shift baseline $F_{int}$ is calculated as the mean median frequency (MMF) at the beginning of the frequency trajectory. The downshift of the median frequency can be estimated by observing the trajectory. 
Then, $F_{th}$ is set as $1.25$ Hz for poor posture sitting to fit the data collected in Experiment 1 in the designed algorithm.

The proposed detection algorithm selects the WM bound variation rate $\Delta_r$ as $0.0083$ for poor posture sitting by fitting the data collected in Experiment 1. Our future work will focus on systematically selecting  WM bound to ensure that the proposed WM-based detection algorithm can achieve good performance in the presence of knowledge of measurement noises and human variations. 

The WM value threshold $WM_{th}$ is triggered by a WM value which is lower than $-0.5$. Given that a WM value can reveal the trend of median frequency trajectory, this threshold is introduced because a WM value lower than $-0.5$ ensures that the frequency trajectory is keeping a decreasing trend whilst allowing the existence of the fluctuations of the trajectory caused by human variations and measurement noises.

%
%
%


\section{Experiment and Signal Processing}

As indicated in Introduction, two set of experiments were conducted. Experiment 1 is served as a training set to train the tuning parameters of the WM-based muscle fatigue detection. Experiment 2 is served as the testing set to test the performance of the proposed algorithm using new subjects and the expertise of the physiotherapist. Two experiments had the same set-up,  procedure, signal processing technique, and requirement. In Experiment 2, an experienced physiotherapist was recruited to evaluate muscle condition every three minutes from the beginning of each trial. Given that muscle fatigue is when the maximum voluntary contraction force is induced \cite{Wang2020}, the physiotherapist used a $3$-point scale ($0$-no stiffness, $1$-moderate stiffness, $2$-hard stiffness) to assess muscle fatigue. 

\subsection{Subjects}

\noindent \underline {Experiment 1}: Sixteen healthy male subjects and one female subject were recruited for the experiment. For two of them, the data were invalid because the looseness of sEMG sensors during the experiment. The characteristic information of the remaining fifteen subjects is shown in Table \ref{tab:subjects}. 

\begin{table}[b]
\vspace{-0.3cm}
\caption{Characteristic information of subjects\label{tab:subjects}}
\begin{center}
\begin{tabular}{c|cccc}
\hline
 & Maximum & Minimum & Median & Mean $\pm$ SD \\
\hline
Age (y) & 29 & 20 & 23 & 23.93 $\pm$ 2.55 \\
Height (cm) & 195 & 168 & 180 & 179.87 $\pm$ 7.38 \\
Weight (kg) & 115 & 66 & 77.5 & 82.10 $\pm$ 14.65 \\
\hline
\end{tabular}
\end{center}
\end{table}

\noindent \underline {Experiment 2}: 6 healthy male subjects were recruited in the experiment  (Participant 16 - Participant 21). In addition, an experienced physiotherapist was recruited to evaluate muscle condition every three minutes from the beginning of each trial. 

In both experiments, informed written consent was obtained from each subject before the experiment. The project was approved by the Human Research Ethics Committee of the University of Melbourne with ID \#1954575.

\subsection{Experimental Setup}

\begin{figure}[b]
    \vspace{-0.4cm}
    \centering
    \begin{subfigure}[h]{0.13\textwidth}
        \centering
        \includegraphics[width=\linewidth]{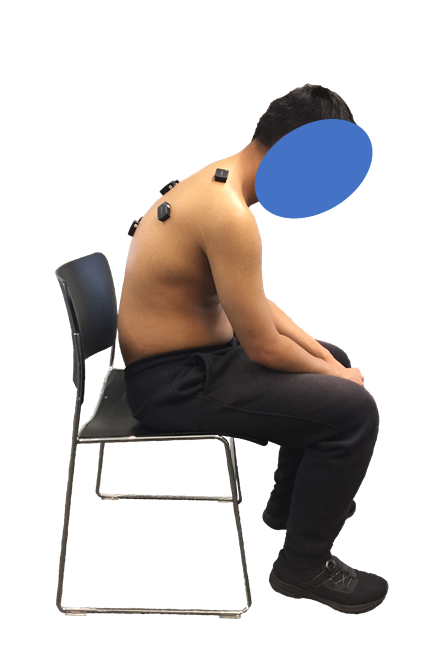}
        \caption{}
    \end{subfigure}
    \begin{subfigure}[h]{0.13\textwidth}
        \centering
        \includegraphics[width=\linewidth]{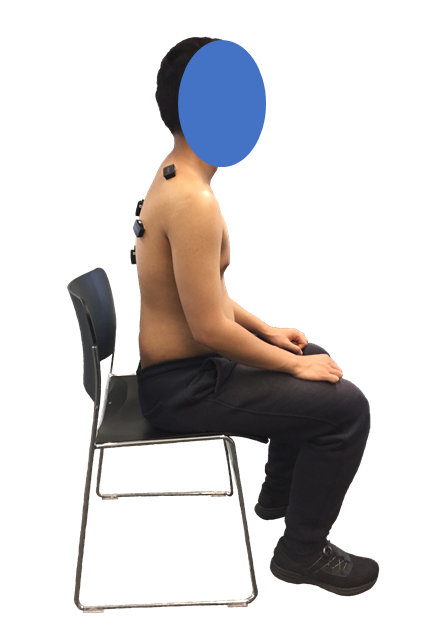}
        \caption{}
    \end{subfigure}
    \begin{subfigure}[h]{0.21\textwidth}
        \centering
        \includegraphics[width=\linewidth]{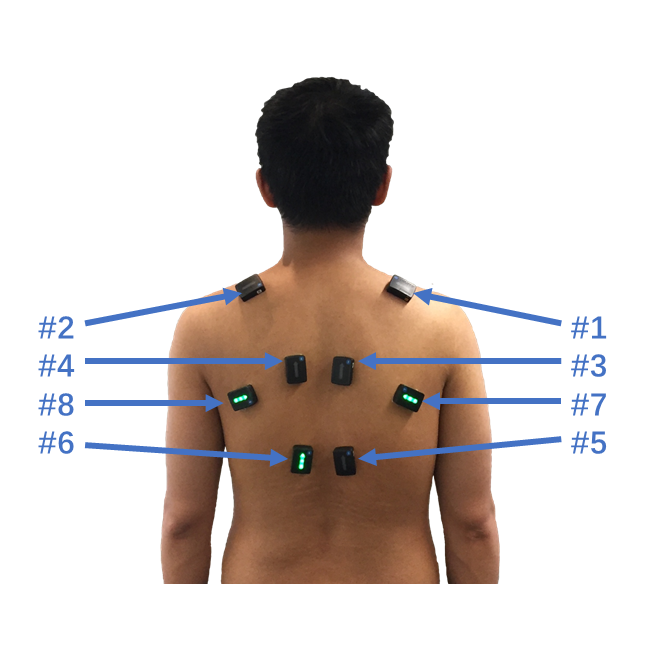}
        \caption{}
    \end{subfigure}
    \caption{\centering Experimental setup of static poses. (a) poor posture. (b) natural posture. (c) placement of sEMG sensors.}
    \label{fig:poor posture}
\end{figure}

In both Experiment 1 and Experiment 2, the subjects were required to keep sitting in a poor posture, which is shown in Fig. \ref{fig:poor posture}(a) for $15$ minutes. The poor posture is a forward head and rounded shoulder posture comparing to the natural posture as Fig. \ref{fig:poor posture}(b) shows. During the experiment, eight sEMG sensors (DELSYS Trigno Biofeedback System, DELSYS Inc., USA) were placed on upper back muscles demonstrated in Fig. \ref{fig:poor posture}(c). They are designed to record sEMG signals from right upper trapezius (sensor \#1), left upper trapezius (sensor \#2), right middle trapezius (sensor \#3), left middle trapezius (sensor \#4), right lower trapezius (sensor \#5), left lower trapezius (sensor \#6), right infraspinatus (sensor \#7) and left infraspinatus (sensor \#8) respectively.

\vspace{-0.35cm}
\subsection{sEMG Signal Processing} 
\begin{figure}[t]
    \centering
    \includegraphics[width=0.8\linewidth]{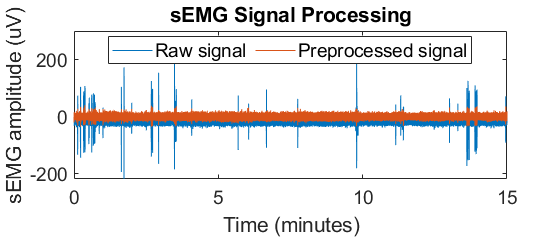}  
    \caption{\centering sEMG signal pre-processing.}
    \label{fig:signalprocessing}
    \vspace{-0.4cm}
\end{figure}

The raw sEMG signal is sampled with $2148$ Hz during data collection. The outliers which are defined as the values outside three standard deviations from the mean are removed. Then the sEMG signals are filtered by the $6^{th}$ order Butterworth band-pass filter with the effective frequency range of sEMG signals $10$ Hz to $500$ Hz, and a $2^{nd}$ order Butterworth band-stop filter with cutoff frequencies $49$ Hz and $51$ Hz to remove the power frequency noise. Fig. \ref{fig:signalprocessing} demonstrates the original raw sEMG signal and the pre-processed signal.

Wavelet analysis is applied to analyse the pre-processed sEMG signals, and decompose the signals into $64$ frequency bands with a frequency interval $16.78$ Hz. Specifically, Daubechies wavelet (db14), Symlet wavelet (sym7) and Coiflets wavelet (coif2) are used. Then, the analysis of variance (ANOVA) function is used to obtain p-values of the median frequencies of sEMG signals at the $1^{st}$, $7^{th}$ and $14^{th}$ minute, representing the beginning, middle and end of the duration of sitting in a poor posture. A p-value less than $0.05$ is considered statistically significant. Such information can be employed to figure out which muscle is sensitive to the muscle fatigue during this static pose when analysing the data. As a result, the frequency band \#5 of the left upper trapezius (sensor \#2) has a p-value 0.044 when analysing with the Daubechies wavelet (db14), which is employed for muscle fatigue detection algorithm development in this study.


\section{Results}

\subsection{Proposed WM-Based Algorithm Tuned from Data Collected from Experiment 1}

\begin{figure}[b]
    \vspace{-0.2cm}
    \centering
    \begin{subfigure}[h]{0.235\textwidth}
        \centering
        \includegraphics[width=\linewidth]{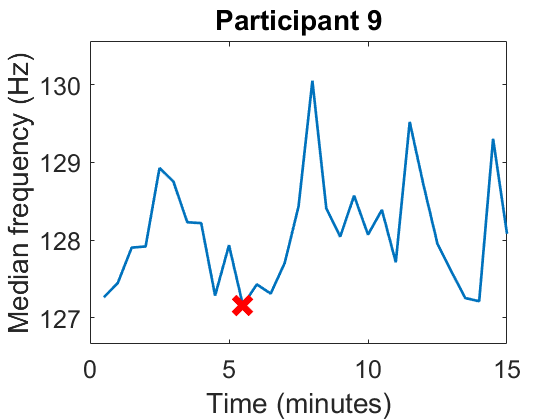} 
        \caption{}
    \end{subfigure}
    \begin{subfigure}[h]{0.235\textwidth}
        \centering
        \includegraphics[width=1.0\linewidth]{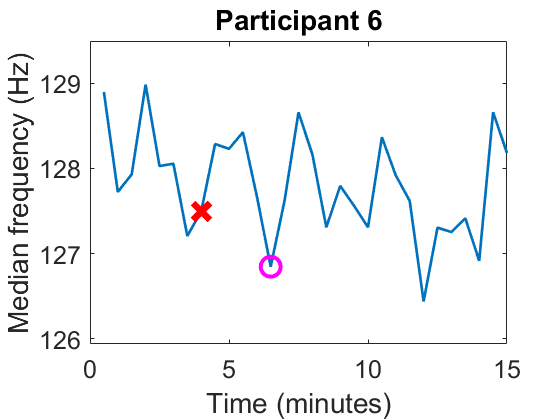}        
        \caption{}
    \end{subfigure}
    \framebox{\parbox{3.1in}{
    \begin{subfigure}[h]{0.45\textwidth}
        \centering
        \includegraphics[width=1.0\linewidth]{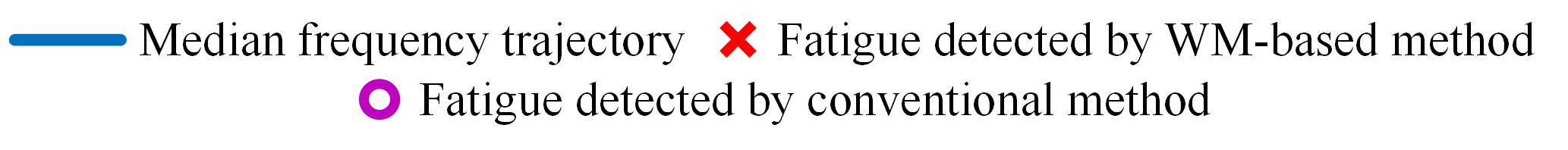}        
    \end{subfigure}   
    }}
    \caption{\centering Representative examples of fatigue detection results. (a) Case 1. (b) Case 2.}
    \label{fig:results1}
\end{figure}

The developed WM-based fatigue detection algorithm was extracted from and applied to total $15$ subjects during their sitting in a poor posture. The proposed method is compared to the conventional threshold-based method, where the threshold is obtained as median frequency decline of $1.25$ Hz.
%
%
Case 1 shows that the WM-based method is more sensitive to successfully detect muscle fatigue for static poses compared with the traditional method.
%
%
%
%
%
The result in Table \ref{tab:comparison} shows that the WM-based method can detect muscle fatigue of $14$ subjects while the conventional method can only detect $6$ subjects. Fig. \ref{fig:results1}(a) shows a representative example that muscle fatigue is detected by the WM-based method at $5.5$ minutes but cannot be detected by the conventional method.


Case 2 is defined for the subjects in the experiment whose muscle fatigue can be identified from both WM-based method and conventional threshold-based method. We use the time when fatigue is detected as the performance index to check the effectiveness. For convenience of notation, $T_{WM}^t$ and $T_{Th}^t$ are denoted for the time instant when muscle fatigue is detected. If $T_{WM}^t<T_{Th}^t$, the index $P_c=1$; and $P_c=-1$ when $T_{WM}^t \geq T_{Th}^t$.

As shown in Table \ref{tab:comparison}, the WM-based method can detect muscle fatigue much earlier.  Fig. \ref{fig:results1}(b) presents a representative example that the WM-based method detected the muscle fatigue at $4$ minutes comparing with the conventional method at $6.5$ minutes. 
 

%
%

\subsection{Validation through Experiment 2}

\begin{figure}[t]
    \centering
    \begin{subfigure}[h]{0.235\textwidth}
        \centering
        \includegraphics[width=\linewidth]{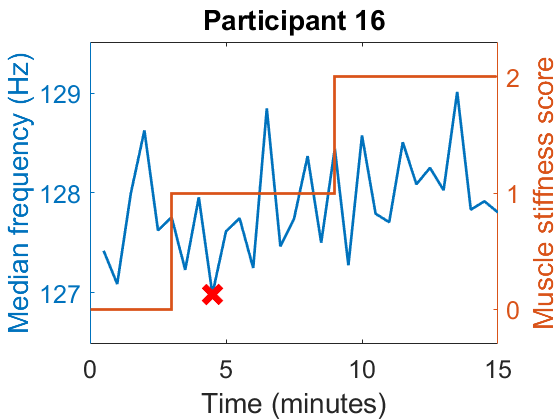}
    \end{subfigure}
    \begin{subfigure}[h]{0.235\textwidth}
        \centering
        \includegraphics[width=\linewidth]{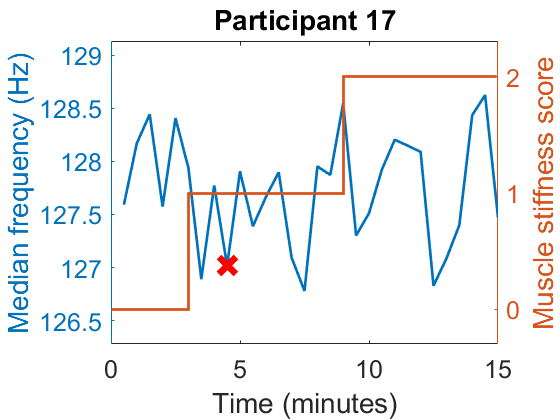}
    \end{subfigure}    
    \begin{subfigure}[h]{0.235\textwidth}
        \centering
        \includegraphics[width=\linewidth]{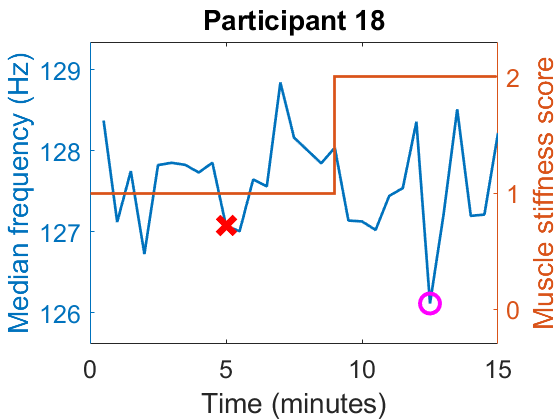}
    \end{subfigure}    
    \begin{subfigure}[h]{0.235\textwidth}
        \centering
        \includegraphics[width=\linewidth]{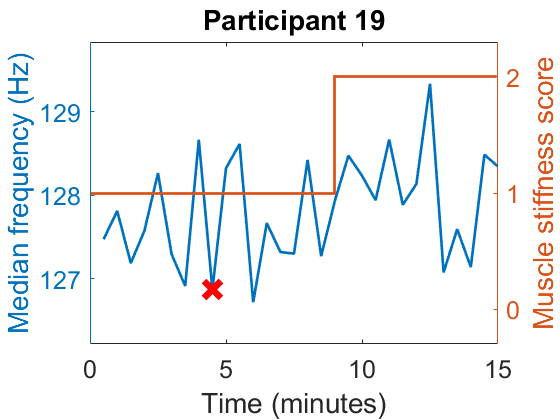}
    \end{subfigure}    
    \begin{subfigure}[h]{0.235\textwidth}
        \centering
        \includegraphics[width=\linewidth]{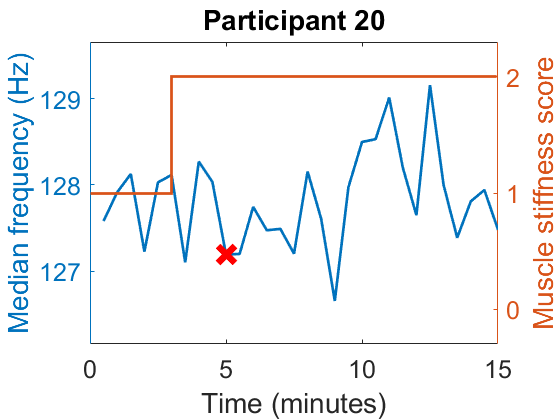}
    \end{subfigure}    
    \begin{subfigure}[h]{0.235\textwidth}
        \centering
        \includegraphics[width=\linewidth]{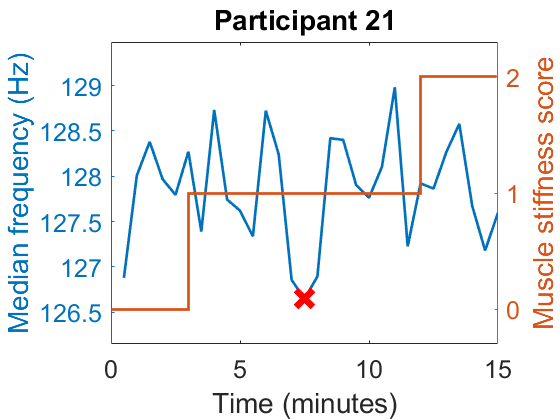}
    \end{subfigure}
    \framebox{\parbox{3.1in}{
    \begin{subfigure}[h]{0.45\textwidth}
        \centering
        \includegraphics[width=1.0\linewidth]{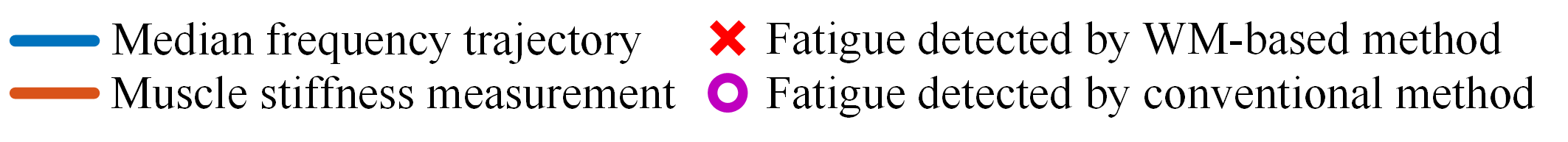}        
    \end{subfigure}   
    }}    
    \caption{\centering Validation results for Experiment 2}
    \label{fig:validexp}
    \vspace{-0.4cm}
\end{figure}

The WM-based algorithm is able to detect muscle fatigue for all $6$ subjects in the algorithm verification experiment which is shown in Fig. \ref{fig:validexp}. In the figure, the step signal is the stiffness score obtained by an experienced physiotherapist, and used as a baseline to indicate the muscle fatigue. 
It can be seen from Fig. \ref{fig:validexp} that the fatigue can be efficiently captured by the WM-based method before the hard stiffness of muscle condition happens for $5$ subjects, which indicates the WM-based method can provide leading time detection of the muscle fatigue. Table \ref{tab:comparison} presents the comparison results of the two methods on detecting muscle fatigue, it can be seen from Table \ref{tab:comparison} that the proposed algorithm has a leading detection time compared to the conventional threshold-based method.

\begin{table}[b]
\vspace{-0.3cm}
\caption{\centering Comparison results between the two methods\label{tab:comparison}}
\begin{center}
\begin{tabular}{cccc}
\hline
Experiment & Categories & \begin{tabular}{@{}c@{}}WM-based\\method\end{tabular} & \begin{tabular}{@{}c@{}}Conventional\\method\end{tabular} \\
\hline
\multirow{2}{*}{Experiment 1} & Case 1 & 14/15 (93.3\%) & 6/15 (40\%) \\
 & Case 2 $P_c=1$ & 5/6 (83.3\%) & - \\
\hline
\multirow{2}{*}{Experiment 2} & Case 1 & 6/6 (100\%) & 1/6 (16.7\%) \\
 & Case 2 $P_c=1$ & 1/1 (100\%) & -\\
\hline
\end{tabular}
\end{center}
\end{table}

\section{Discussion}
The comparison results show that both the WM-based method and the conventional method are able to detect muscle fatigue for a number of subjects. However, the WM-based method can successfully detect $8$ more subjects out of $15$ in Experiment 1 and $5$ more subjects out of $6$ in Experiment 2 than the convention method. These results indicate that the WM-based method is more robust in detecting muscle fatigue during static poses though there are measurement noises and human variations at all median frequency trajectories.

At the same time, the WM-based method can detect muscle fatigue much earlier than the conventional method.  The early detection is important for the prevention and prediction analysis because it allows time to prevent muscle injury caused by chronic and excessive muscle fatigue. 


It is also noted that there is one subject whose muscle fatigue cannot be detected by neither the WM-based method nor the conventional method in Experiment 1. This might come from large human variations. In particular, this subject felt more comfortable in the poor posture, suggesting that the definition of poor posture is subject-related. How to define the personalized poor posture is very challenging. Our future work will focus on it.

There are also some limitations in this study. Firstly, the muscle fatigue during natural posture is not measured. The experiment set-up was based on the experience of the physiotherapist, indicating that subjects will fatigue when they are sitting in a given (well-known) poor posture for $15$ minutes. The data from male subjects is used for convenience of taking measurements, which leads to the limited validity of the proposed fatigue detection algorithm. 


\section{Conclusion}
A novel fatigue detection algorithm for upper-back muscle fatigue during a short-duration poor posture was proposed, based on the population trend observed in the muscle fatigue procedure in term of median frequency trend of sEMG signals. 
The concept of weak monotonicity (WM), which is a robust population trend, is thus utilized to detect the muscle fatigue in the presence of measurement noises and human variations. 
The experimental results show that the WM-based detection algorithm is more sensitive in successfully detecting muscle fatigue with the possibility of early detection.

\addtolength{\textheight}{-12cm}   









\bibliographystyle{IEEEtran}
\bibliography{root}

\end{document}